\documentclass[preprint,aps,titlepage,twocolumn,tightenlines]{revtex4}%
\usepackage{amssymb}
\usepackage{amsfonts}
\usepackage{amsmath}
\usepackage{graphicx}
\usepackage{hyperref}%
\providecommand{\U}[1]{\protect\rule{.1in}{.1in}}

\begin{document}
\title[Line Segments]{On the fields due to line segments}
\author{T. S. Van Kortryk}
\affiliation{120 Payne Street, Paris, MO65275}

\begin{abstract}
The remarkable geometries of ellipsoidal equipotentials and their associated
gradient fields, as produced by uniformly charged or current carrying
straight-line segments, are discussed at an elementary level, motivated by
recent treatments intended for introductory physics classes. \ Some effort is
made to put the results into a broader conceptual and historical context.
\ The equipotentials and vector fields were first obtained for the
electrostatic problem by George Green in his famous 1828 essay. \ Related
problems often appeared on the Mathematical Tripos examinations given at the
University of Cambridge, and their solutions were widely disseminated by
William Thomson (Lord Kelvin), Peter Guthrie Tait, and Edward Routh during the
last half of the 19$^{\text{th}}$ century.

\end{abstract}
\startpage{1}
\endpage{ }
\maketitle

\section{Introduction}

There are a number of problems in electromagnetism where coordinates
\emph{centered on the observation point} simplify the integrations required to
obtain either the potentials or the fields.
\ \href{http://posner.library.cmu.edu/Posner/books/pages.cgi?call=537_M46T_1873_VOL._1&layout=vol0/part0/copy0&file=0110}{The
standard example} is to show there is no electric field at any point inside of
a uniformly charged spherical shell by choosing spherical polar coordinates
centered on the point in question \cite{Maxwell,SideRemark1}. \ For another
illustration, the electric \emph{potential} on the rim of a uniformly charged
disk is most readily computed using such coordinates, as shown in Purcell and
Morin \cite{Purcell}, page 70, Eqn (2.30). \ 

The magnetic field along the axis of a finite solenoid is also very easily
obtained using polar coordinates about the observation point (again see
\cite{Purcell}, page 300). \ But Purcell and Morin do \emph{not} compute
$\overrightarrow{E}$ for a uniformly charged, \emph{finite length}, straight
line segment using this method. \ Nor does the author of any other textbook in
\emph{common} use today, as far as I can tell, including Griffiths
\cite{Griffiths} and Jackson \cite{Jackson}. \ Even in his treatise
\cite{Maxwell} Maxwell does not solve the finite line segment problem using
coordinates centered on the observation point.

\section{A Charge Segment}

Recently, however, Zuo \cite{Zuo} has presented a derivation of the electric
field produced by the straight-line segment, through the use of coordinates
centered on the observation point. (The reader is encouraged to read
\cite{Zuo} before proceeding.) \ In terms of the coordinates used by Zuo, the
electric field integral reduces to
\begin{equation}
\frac{1}{y}\int\widehat{n}\left(  \theta\right)  ~d\theta\label{EIntegral}%
\end{equation}
where $\widehat{n}\left(  \theta\right)  $ is the local normal to a circle of
fixed radius $y$ whose center is the observation point. \ While Zuo might have
found a novel way to do the calculation, it seems highly unlikely that there
is no precedent given that this particular problem must have been studied by
many people \cite{Rowley} during the 140+ years since Maxwell's treatise first
appeared \cite{SideRemark2}. \ 

In fact, this calculational method was already known to work very well for the
line segment. \ It was discussed and widely disseminated by Edward Routh in
the 19$^{\text{th}}$ century, and it was probably familiar to almost every
student at that time who took the famous
\href{http://en.wikipedia.org/wiki/Cambridge_Mathematical_Tripos}{Mathematical
Tripos examinations} at the University of Cambridge during
\href{http://books.google.com/books?id=tRnwfbg_O1gC&pg=PA229&dq=edward+routh+and+the+art+of+good+coaching&hl=en&sa=X&ei=0h87VLa-Bcjm8QGYioGYCA&ved=0CDAQ6AEwAA#v=onepage&q=5.1.%20Training%20and%20Research&f=false#v=onepage&q&f=false}{Routh's
unsurpassed coaching of students} \cite{Warwick} for those examinations
\cite{SideRemark3}. \ 

More specifically, an\ exact solution of the line segment problem, making use
of calculus and coordinates centered on the observation point, was published
more than 120 years ago as the very first example on
\href{https://play.google.com/books/reader2?id=YvFJAAAAMAAJ&printsec=frontcover&output=reader&hl=en&pg=GBS.PA4}{pages
4-6, volume II}, of Routh's once widely-read
\href{http://www.amazon.com/s/ref=nb_sb_noss?url=search-alias%3Daps&field-keywords=%29.+A+Treatise+on+Analytical+Statics+with+Numerous+Examples+Routh&rh=i%3Aaps%2Ck%3A%29.+A+Treatise+on+Analytical+Statics+with+Numerous+Examples+Routh}{books
on analytical statics} \cite{Routh}. \ The solution includes two clearly drawn
diagrams. \ Although Routh discussed the problem in the context of Newtonian
gravity, the mathematics is exactly the same in that context as it is for the
electrostatic problem. \ It is immediately evident that Routh's and Zuo's
methods are identical.

Even earlier, William Thomson (Lord Kelvin) and Peter Guthrie Tait had
published a solution to the same gravitational problem from the same point of
view using only geometrical reasoning without calculus (as if following
Newton's lead \cite{SideRemark1}) but arriving at the same results
\cite{KelvinTait}
(\href{http://babel.hathitrust.org/cgi/pt?id=uc1.b3622012;view=1up;seq=58}{Volume
II, Section 481, pages 26-28}) while making use of similar diagrams. \ More
recently (i.e. \emph{only} 60 or 50 years ago) the electric field was
discussed from the same perspective in \cite{Durand53}, pp 50-51, as well as
in \cite{Durand64}, volume I, pp 155-156, again with similar diagrams
\cite{Puzzling}.

Still, even though the line segment problem has been solved several times
before by almost exactly the same method, I would agree with Zuo that
\emph{many people }$\boldsymbol{today}$\emph{\ do not know} either the method
or that it works so well for this problem. \ In any case, this is a remarkably
simple continuous charge distribution where Gauss' law does \emph{not}
trivially give the answer, but nevertheless the integral to obtain the
electric field from Coulomb's law \emph{is} trivial to evaluate, in special
coordinates, and therefore more easily computed than even the potential.

\subsection{Green's potential}

Of course the electric potential can also be computed using the same
coordinates. \ But the integral for the potential does \emph{not} reduce
simply to $\int d\theta$ as one might \emph{naively} expect given the form of
the electric field integral in (\ref{EIntegral}). \ In contrast, the integral
required for the potential in those same coordinates is%
\begin{equation}
\int\frac{d\theta}{\cos\theta}=\ln\left(  \frac{1+\sin\theta}{\cos\theta
}\right)  \label{VIntegral}%
\end{equation}
Referring to Zuo's first figure, this immediately gives the result
($1/k=4\pi\varepsilon_{0}$)%
\begin{equation}
V=k\lambda\ln\left(  \frac{b+r_{b}}{a+r_{a}}\right)  \label{V}%
\end{equation}
where $r_{a}$ and $r_{b}$ are the distances from the observation point to the
left and right ends of the line segment, located on the $x$-axis at $a$ and
$b$, respectively. \ With a little algebra \cite{FootnoteA} this potential can
be rewritten as%
\begin{equation}
V\left(  s\right)  =k\lambda\ln\left(  \frac{s+L}{s-L}\right)  \text{
},\ \ \ s=r_{a}+r_{b} \label{VEllipsoid}%
\end{equation}
where $L$ is the length of the line segment ($L=b-a$ in Zuo's coordinates).
\ In this form, the equipotentials, which are given by constant $r_{a}+r_{b}$,
are clearly just prolate ellipsoids of revolution about the axis of the line
segment. \ 

So far as I have been able to determine, the result (\ref{VEllipsoid}) first
appears under Article 12 in the brilliant 1828 essay by George Green
\cite{Green} (see
\href{http://books.google.com/books?id=u4JsAAAAMAAJ&printsec=frontcover&dq=george+green&hl=en&sa=X&ei=jO86VLegCM-L8gGswYHQBA&ved=0CCUQ6AEwAQ#v=onepage&q=line%20uniformly%20covered&f=false}{pp
68-69} in \cite{Ferrers}). \ Commenting on Green's essay several decades
later, in 1870, N M Ferrers aptly summarized the situation in an
\href{http://books.google.com/books?id=u4JsAAAAMAAJ&printsec=frontcover&dq=george+green&hl=en&sa=X&ei=jO86VLegCM-L8gGswYHQBA&ved=0CCUQ6AEwAQ#v=onepage&q=Appendix&f=false}{Appendix
to Green's collected papers} (p 329 in \cite{Ferrers}):

\begin{quote}
\textsf{In the case of a straight line uniformly covered with electricity ...
Denoting the extremities of the straight line by }$S,H$\textsf{, we know that
the attraction of the line on }$p$\textsf{ ... may be replaced by that of a
circular arc of which }$p$\textsf{ is the centre ... \ Hence the direction of
the resultant attraction bisects the angle }$SpH$\textsf{, and the
equipotential surface is a prolate spheroid of which }$S,H$\textsf{ are the
foci.}
\end{quote}

\noindent Thus it would seem the essential features of both
$\overrightarrow{E}$ and $V$ for the uniformly charged line segment were
understood and fully appreciated as a consequence of Green's work
\cite{SideRemark4}.

Today the role played by ellipsoidal equipotentials for the charged line
segment is well-known
\cite{Durand53,Durand64,Smythe,Chandrasekhar,Rowley,Zangwill}. \ In my
opinion, most physicists would agree that the geometry of these ellipsoids is
the \textquotedblleft hidden symmetry\textquotedblright\ that underlies the
line segment problem \cite{SideRemark5}.

It is also well-known that the normals to an ellipse will always bisect the
angle formed by the $r_{a}$ and $r_{b}$ lines \cite{FootnoteB}. \ Thus the
direction of the electric field for the uniformly charged line segment will
also bisect this angle, since $\overrightarrow{E}$ is always normal to
equipotential surfaces \cite{FootnoteC}. \ This agrees with Routh's and Zuo's
conclusion based on the explicit integral (\ref{EIntegral}). \ But in
consideration of the well-known geometry of an ellipse, and the early work of
Green, it is definitely \emph{not} appropriate to say that a calculation using
coordinates centered on the observation point (such as that by Routh or Zuo)
is either the first or the only way the direction of the total electric field
for this charge configuration can be graphically defined.\ \ 

On the other hand a calculation based on a perspective from the observation
point is technically sweet, and the resulting form for the magnitude of
$\overrightarrow{E}$\ has some interesting features. \ To shed more light on
those features, it is useful to compare Routh's and Zuo's result for the form
of $\left\vert \overrightarrow{E}\right\vert $ to that obtained directly from
the potential as given by (\ref{VEllipsoid}).

\subsection{Various results for $\protect\overrightarrow{E}$}

Given the coordinate-independent expression for the potential,
(\ref{VEllipsoid}), the electric field may be obtained by elementary vector
calculus, without reference to explicit coordinates. \ To achieve this let
$\overrightarrow{r_{a}}$ and $\overrightarrow{r_{a}}$\ be vectors from the $a$
and $b$ ends of the line segment to the observation point, let
$\overrightarrow{r}$ be the vector from the \emph{center} of the segment to
the observation point, and let $\overrightarrow{L}$ be the vector from point
$a$ to point $b$. \ Then%
\begin{equation}
\overrightarrow{r_{a}}=\overrightarrow{r}+\frac{1}{2}\overrightarrow{L}%
\ ,\ \ \ \overrightarrow{r_{b}}=\overrightarrow{r}-\frac{1}{2}%
\overrightarrow{L} \label{VectorDefinitions}%
\end{equation}
and
\[
\overrightarrow{\nabla}r_{a,b}=\overrightarrow{\nabla}\sqrt{\left(
\overrightarrow{r}\pm\frac{1}{2}\overrightarrow{L}\right)  \cdot\left(
\overrightarrow{r}\pm\frac{1}{2}\overrightarrow{L}\right)  }=\frac{\left(
\overrightarrow{r}\pm\frac{1}{2}\overrightarrow{L}\right)  }{\sqrt{\cdots}}%
\]
so the gradients of $r_{a}$\ and $r_{b}$\ are simply \emph{unit}
\emph{vectors}.
\begin{equation}
\overrightarrow{\nabla}r_{a}=\frac{\overrightarrow{r_{a}}}{r_{a}}%
\equiv\widehat{r_{a}}\ ,\ \ \ \overrightarrow{\nabla}r_{b}=\frac
{\overrightarrow{r_{b}}}{r_{b}}\equiv\widehat{r_{b}}%
\end{equation}
From these elementary facts, and (\ref{VEllipsoid}), $\overrightarrow{E}%
\left(  \overrightarrow{r}\right)  $\ may be computed in the usual way.%
\begin{align}
\overrightarrow{E}\left(  \overrightarrow{r}\right)   &
=-\overrightarrow{\nabla}V\left(  s\right)  =-\frac{dV\left(  s\right)  }%
{ds}~\overrightarrow{\nabla}s\nonumber\\
&  =-\frac{dV\left(  s\right)  }{ds}~\left(  \overrightarrow{\nabla}%
r_{a}+\overrightarrow{\nabla}r_{b}\right) \nonumber\\
&  =-\frac{dV\left(  s\right)  }{ds}~\left(  \widehat{r_{a}}+\widehat{r_{b}%
}\right)  \label{EasGradientV}%
\end{align}
That is to say, the direction of $\overrightarrow{E}\left(  \overrightarrow{r}%
\right)  $ is given just by the arithmetic average of the unit vectors
$\widehat{r_{a}}$ and $\widehat{r_{b}}$. \ But these unit vectors form the
equal-length sides of an \emph{isosceles} triangle, and their vector sum
therefore bisects the angle between them \cite{Durand64,Rowley}. \ This
establishes yet again that $\overrightarrow{E}$ bisects the angle $\theta
_{ab}$ between $\overrightarrow{r_{a}}$ and $\overrightarrow{r_{b}}$.

Moreover, the magnitude of the electric field is now explicitly given in terms
of $s$ and $\theta_{ab}=\arccos\left(  \widehat{r_{a}}\cdot\widehat{r_{b}%
}\right)  $, upon using
\begin{equation}
-\frac{dV\left(  s\right)  }{ds}=\frac{2k\lambda L}{s^{2}-L^{2}} \label{V'}%
\end{equation}
Consequently I obtain the magnitude of the electric field in a different form
than that exhibited by Routh and Zuo.%
\begin{align}
\left\vert \overrightarrow{E}\left(  \overrightarrow{r}\right)  \right\vert
&  =\left\vert \frac{dV\left(  s\right)  }{ds}\right\vert \sqrt{\left(
\widehat{r_{a}}+\widehat{r_{b}}\right)  \cdot\left(  \widehat{r_{a}%
}+\widehat{r_{b}}\right)  }\nonumber\\
&  =\frac{2kL\left\vert \lambda\right\vert }{s^{2}-L^{2}}~\sqrt
{2+2\widehat{r_{a}}\cdot\widehat{r_{b}}}\nonumber\\
&  =\frac{4kL\left\vert \lambda\cos\left(  \theta_{ab}/2\right)  \right\vert
}{s^{2}-L^{2}} \label{EMagnitude}%
\end{align}
Now, this too is a well-known result (e.g. see
\cite{Routh,Durand53,Durand64,Smythe,Chandrasekhar,Rowley}). \ The $\left\vert
\frac{dV\left(  s\right)  }{ds}\right\vert $ factor in $\left\vert
\overrightarrow{E}\left(  \overrightarrow{r}\right)  \right\vert $ is constant
on any of the equipotential ellipsoids, but the angle-dependent factor
$\cos\left(  \theta_{ab}/2\right)  $ varies, in general. \ Note that $s>L$ for
all those observation points that do not lie on the line segment itself.

Also note the transparent behavior of $\overrightarrow{E}$ as given by
(\ref{EMagnitude}) in some situations. \ For example, far away from the the
line segment, $r\gg L$, so $s^{2}-L^{2}\approx s^{2}\approx4r^{2}$ and
$\cos\left(  \theta_{ab}/2\right)  \approx\cos\left(  0\right)  =1$. \ Thus
the field looks like a point charge, $\left\vert \overrightarrow{E}\left(
\overrightarrow{r}\right)  \right\vert \approx kL\left\vert \lambda\right\vert
/r^{2}$, as expected. \ Also, for points $\overrightarrow{r}=\pm
s~\widehat{L}$ with $s>L$, i.e. collinear with the segment but outside of it,
the field reduces to a well-known form. \ For such points, $\cos\left(
\theta_{ab}/2\right)  =\cos\left(  0\right)  =1$ and $s=2r$.

While (\ref{EMagnitude}) is a simple result for $\left\vert \overrightarrow{E}%
\left(  \overrightarrow{r}\right)  \right\vert $, it's behavior is not always
completely transparent, and it is not obviously equivalent to the form given
by Routh and Zuo. \ For instance, in the limit where the observation point
transversely approaches some interior point on the straight line joining $a$
and $b$, the charged segment should be indistinguishable from an infinitely
long straight line charge. \ That is to say, it should be true that
$y\left\vert \overrightarrow{E}\left(  \overrightarrow{r}\right)  \right\vert
\underset{y{\footnotesize \rightarrow0}}{\longrightarrow}2k\lambda$ where $y$
is the \textquotedblleft$\perp$\ distance\textquotedblright\ from the
observation point to the line of charge. \ On the other hand, as interior
points are approached, $\lim_{{\footnotesize y\rightarrow0}}\cos\left(
\theta_{ab}/2\right)  =\cos\left(  \pi/2\right)  =0$, so the $s^{2}-L^{2}$
denominator in (\ref{EMagnitude}) better have a double zero and vanish like
$y\cos\left(  \theta_{ab}/2\right)  $ to obtain the correct limit. \ It does.

Although a coordinate-free proof from first principles might be challenging
for an inexperienced student, it is nevertheless true that \cite{FootnoteD}%
\begin{equation}
\left(  s^{2}-L^{2}\right)  \left\vert \tan\left(  \theta_{ab}/2\right)
\right\vert =2hL \label{Identity}%
\end{equation}
where $h\geq0$ is the $\perp$ distance from the infinite straight line
containing the segment to the point in question on the ellipse. \ Thus the
result (\ref{EMagnitude}) may also be written as%
\begin{equation}
\left\vert \overrightarrow{E}\left(  \overrightarrow{r}\right)  \right\vert
=\frac{2k\left\vert \lambda\sin\left(  \theta_{ab}/2\right)  \right\vert }{h}
\label{ZuoForm}%
\end{equation}
This is the form obtained by Routh and Zuo directly from integration performed
from the perspective of the observation point. \ The results (\ref{EMagnitude}%
)\ and (\ref{ZuoForm})\ are therefore completely equivalent expressions for
the same electric field. \ Still, because it can be somewhat painful to
establish (\ref{Identity}), and because the standard treatment of this problem
involves first finding the potential and \emph{then} finding
$\overrightarrow{E}$, this latter form for $\left\vert \overrightarrow{E}%
\right\vert $ is not the one most likely to be found in intermediate or more
advanced texts as routinely used today.

The result (\ref{ZuoForm}) has some features that nicely complement those of
(\ref{EMagnitude}), and vice versa. \ As one rather obvious feature,
(\ref{ZuoForm}) consists of a simple geometrical factor multiplying the field
that would be produced by an infinitely long uniformly charged straight line
(from $-\infty$ to $+\infty$). \ That is,
\begin{equation}
\left\vert \overrightarrow{E}\left(  \overrightarrow{r}\right)  \right\vert
=\left\vert \overrightarrow{E}_{\infty}\sin\left(  \frac{\theta_{ab}}%
{2}\right)  \right\vert \ ,\ \ \ \left\vert \overrightarrow{E}_{\infty
}\right\vert =\left\vert \frac{2k\lambda}{h}\right\vert \label{FactoredForm}%
\end{equation}
where again $h$ is the $\perp$\ distance from the observation point to the
infinite line containing the charged segment. \ The $\sin\left(  \theta
_{ab}/2\right)  $ geometrical factor brings to mind some other well-known
examples of static fields \cite{FootnoteE}. \ The general form (but not the
specific dependence on the angles) follows just from elementary dimensional analysis.

As a consequence of (\ref{FactoredForm}), the approach to any point in the
interior of the line segment is now easy to understand, since $\sin\left(
\frac{\theta_{ab}}{2}\right)  \rightarrow\sin\left(  \frac{\pi}{2}\right)
=1$. \ Thus the segment field approaches the infinite line result,
$\overrightarrow{E}_{\infty}$, as $h\rightarrow0$ for any point between $a$
and $b$. \ For this situation, (\ref{FactoredForm})\ is more useful than
(\ref{EMagnitude}).

However, for points $\overrightarrow{r}=\pm s~\widehat{L}$ with $s>L$, it is
necessary to take a careful limit of (\ref{FactoredForm}) to obtain the usual
collinear result since both $\sin\left(  \theta_{ab}/2\right)  =0$ and $h=0$
for such points. \ For this situation, (\ref{EMagnitude})\ is easier to
understand. \ Also, to see the point-like $1/r^{2}$ behavior of the field for
any distant point it is necessary to take a careful limit of
(\ref{FactoredForm}) since $\sin\left(  \frac{\theta_{ab}}{2}\right)
\rightarrow\sin\left(  0\right)  =0$ as $r\rightarrow\infty$. \ Again, for
this situation, (\ref{EMagnitude})\ is more transparent.

A few more remarks are in order before closing this discussion of the electric
field due to a uniformly charged line segment. \ For this problem, as in many
others, knowing the direction of $\overrightarrow{E}$ at any point permits the
complete determination of $\overrightarrow{E}$ just from one (non-zero!)
component. \ In this case it is easy to find the component parallel to the
direction of the segment. \ This component can be found without having to do
\emph{any} integrations.

For instance, if the segment is along the $z$-axis, in cylindrical
coordinates, by azimuthal symmetry $E_{\phi}=0$, and $\left(  E_{\rho}%
,E_{z}\right)  =\left(  \left\vert \overrightarrow{E}\right\vert \sin
\theta_{E},\left\vert \overrightarrow{E}\right\vert \cos\theta_{E}\right)
=\left(  E_{z}\tan\theta_{E},E_{z}\right)  $, where $\theta_{E}$ is the polar
angle of the vector $\overrightarrow{E}$ at the point in question. \ Now,
$E_{z}$ can be determined without actually having to do any integrations ---
the integrations are all eliminated by Dirac deltas. \ To see this note that
$V$ and $\overrightarrow{E}=-\overrightarrow{\nabla}V$ \emph{both} obey
Poisson equations, namely,%
\begin{equation}
\nabla^{2}V=-\frac{1}{\varepsilon_{0}}~\rho\text{ },\text{ \ \ }\nabla
^{2}\overrightarrow{E}=\frac{1}{\varepsilon_{0}}~\overrightarrow{\nabla}\rho
\end{equation}
where $\rho$\ is the local charge density. \ In free space then, without any
boundaries,
\begin{equation}
\overrightarrow{E}\left(  \overrightarrow{r}\right)  =\frac{-1}{4\pi
\varepsilon_{0}}\int\frac{\overrightarrow{\nabla_{s}}\rho\left(
\overrightarrow{s}\right)  }{\left\vert \overrightarrow{r}-\overrightarrow{s}%
\right\vert }~d^{3}s \label{EGradRho}%
\end{equation}
For a uniformly charged segment along the $z$-axis, between $-L/2$ and $L/2$,
say, the charge density is given in terms of Heaviside step functions and
Dirac deltas.%
\begin{equation}
\rho\left(  x,y,z\right)  =\lambda~\theta\left(  \tfrac{L}{2}-z\right)
\theta\left(  z+\tfrac{L}{2}\right)  \delta\left(  x\right)  \delta\left(
y\right)
\end{equation}
Thus the $z$ component of $\overrightarrow{\nabla}\rho$ consists of
three-dimensional Dirac deltas.
\begin{align}
\partial_{z}\rho\left(  x,y,z\right)   &  =\lambda~\delta\left(  z+\tfrac
{L}{2}\right)  \delta\left(  x\right)  \delta\left(  y\right) \nonumber\\
&  -\lambda~\delta\left(  z-\tfrac{L}{2}\right)  \delta\left(  x\right)
\delta\left(  y\right)
\end{align}
So all three integrations in (\ref{EGradRho}) are automatically eliminated for
$E_{z}$. \ The result for any observation point $\overrightarrow{r}$ is
\begin{equation}
E_{z}\left(  \overrightarrow{r}\right)  =\frac{\lambda}{4\pi\varepsilon_{0}%
}\left(  \frac{1}{\left\vert \overrightarrow{r}-\frac{1}{2}L\widehat{z}%
\right\vert }-\frac{1}{\left\vert \overrightarrow{r}+\frac{1}{2}%
L\widehat{z}\right\vert }\right)  \label{Easy}%
\end{equation}
This result along with the direction of $\overrightarrow{E}$\ at any point (as
given by $\widehat{r_{a}}+\widehat{r_{b}}$, say) may be used as an equivalent
alternative to either (\ref{EMagnitude})\ or (\ref{ZuoForm}). \ It is not
surprising that (\ref{Easy}) can also be found in \cite{Routh}%
\ (\href{https://play.google.com/books/reader2?id=YvFJAAAAMAAJ&printsec=frontcover&output=reader&hl=en&pg=GBS.PA4}{see
Volume II, page 5, Eqn(3)}) and in \cite{Durand64} (see Volume I, page 155, Eqn(83)).

\section{A Current Segment}

Straight line segments carrying constant currents also lead to ellipsoidal
equipotentials and associated magnetic vector fields. \ The Biot-Savart law
applied to current $I$ flowing along a directed line segment represented by
the vector $\overrightarrow{L}$ gives a magnetic field due to \emph{only} the
segment as follows:%
\begin{align}
\overrightarrow{B}\left(  \overrightarrow{r}\right)   &  =\widehat{L}%
\times\overrightarrow{C}\left(  \overrightarrow{r}\right) \\
\overrightarrow{C}\left(  \overrightarrow{r}\right)   &  =\frac{\mu_{0}I}%
{4\pi}\int_{-L/2}^{L/2}\frac{\overrightarrow{r}-\ell~\widehat{L}}{\left\vert
\overrightarrow{r}-\ell~\widehat{L}\right\vert ^{3}}~d\ell
\end{align}
where $\overrightarrow{r}$\ is a vector from the center of the segment to the
observation point. \ That is to say, under the replacement $\mu_{0}%
I\rightarrow\lambda/\varepsilon_{0}$ this integral expression for the
auxiliary vector field $\overrightarrow{C}\left(  \overrightarrow{r}\right)  $
is exactly the same as the Coulomb integral for the electric field
$\overrightarrow{E}\left(  \overrightarrow{r}\right)  $ of the previous
uniformly charged segment. \ Consequently $\overrightarrow{C}\left(
\overrightarrow{r}\right)  $ has the same geometric features as that previous
$\overrightarrow{E}\left(  \overrightarrow{r}\right)  $, e.g. the direction
$\widehat{C}\left(  \overrightarrow{r}\right)  $ bisects the angle between
$\overrightarrow{r_{a}}$ and $\overrightarrow{r_{b}}$, where these vectors are
defined as in (\ref{VectorDefinitions}) from the ends of the line segment to
the observation point.

The correspondence between $\overrightarrow{C}$ and the previous charged
segment $\overrightarrow{E}$ also allows us to write%
\begin{align}
\overrightarrow{C}\left(  \overrightarrow{r}\right)   &  =\frac{\mu_{0}I}%
{4\pi}\left(  \frac{2L}{s^{2}-L^{2}}\right)  \left(  \widehat{r_{a}%
}+\widehat{r_{b}}\right) \\
\overrightarrow{B}\left(  \overrightarrow{r}\right)   &  =\frac{\mu_{0}I}%
{4\pi}\left(  \frac{2}{s^{2}-L^{2}}\right)  \overrightarrow{L}\times\left(
\widehat{r_{a}}+\widehat{r_{b}}\right)  \label{BLineSegment}%
\end{align}
Moreover,
\begin{align}
\overrightarrow{C}\left(  \overrightarrow{r}\right)   &
=-\overrightarrow{\nabla}U\left(  s\right)  =-\frac{dU\left(  s\right)  }%
{ds}~\overrightarrow{\nabla}s\\
U\left(  s\right)   &  =\frac{\mu_{0}I}{4\pi}\ln\left(  \frac{s+L}%
{s-L}\right)  \text{ },\ \ \ s=r_{a}+r_{b}%
\end{align}
where $U\left(  s\right)  $ becomes exactly the same as $V\left(  s\right)  $
in (\ref{VEllipsoid}) upon replacing $\mu_{0}I\rightarrow\lambda
/\varepsilon_{0}$.

From these results it follows that $\overrightarrow{B}$ is in the usual form
of a curl,
\begin{equation}
\overrightarrow{B}\left(  \overrightarrow{r}\right)  =\overrightarrow{\nabla
}\times\overrightarrow{A}\left(  \overrightarrow{r}\right)
\end{equation}
where the easily visualized vector potential due to the segment is
\begin{equation}
\overrightarrow{A}\left(  \overrightarrow{r}\right)  =\widehat{L}~U\left(
s\right)
\end{equation}
This $\overrightarrow{A}$ is constant on each ellipsoid of revolution confocal
with $\overrightarrow{L}$.

After evaluating the cross products in (\ref{BLineSegment}) and using the
identity (\ref{Identity}), the result for $\overrightarrow{B}$\ is a simple
geometrical factor multiplying the field $\overrightarrow{B}_{\infty}$ that
would be produced by an infinitely long straight-line current. \ That is,
\begin{align}
\overrightarrow{B}\left(  \overrightarrow{r}\right)   &  =\sin\left(
\tfrac{\vartheta_{b}-\vartheta_{a}}{2}\right)  \sin\left(  \tfrac
{\vartheta_{a}+\vartheta_{b}}{2}\right)  ~\overrightarrow{B}_{\infty}\left(
\overrightarrow{r}\right) \nonumber\\
&  =\frac{1}{2}\left(  \cos\theta_{a}-\cos\theta_{b}\right)
~\overrightarrow{B}_{\infty}\left(  \overrightarrow{r}\right)
\label{BFactoredForm}%
\end{align}%
\begin{equation}
\overrightarrow{B}_{\infty}\left(  \overrightarrow{r}\right)  \equiv\frac
{\mu_{0}I~\widehat{\varphi}}{2\pi h}%
\end{equation}
where $\vartheta_{a}$ and $\vartheta_{b}$\ are the polar angles for
$\overrightarrow{r_{a}}$ and $\overrightarrow{r_{b}}$\ as measured from an
axis along $\overrightarrow{L}$, where $\widehat{\varphi}$ is the azimuthal
unit vector about that axis, and where $h$ is the $\perp$\ distance from the
observation point to that same axis. \ The general form of $\overrightarrow{B}%
$ (but not the specific dependence on the angles) again follows just from
elementary dimensional analysis.

Of course, the magnetic field due to a straight-line segment of current is
treated in many texts (e.g. \cite{Griffiths},\ page 225, Example 5.5, and
\cite{Zangwill}, pp 306-307, Example 10.1), although few if any of these
treatments emphasize parallels between the calculation of $\overrightarrow{B}$
for this situation and the calculation of $\overrightarrow{E}$ for the charged
line segment, as I have done here. \ However, the perspicacious reader of
\cite{Zuo}\ and of the solution for the current segment exhibited in
\cite{Griffiths} will have noticed that both authors use exactly the same
change of variable to evaluate the necessary integral, as well as an identical diagram.

\section{Generalizations}

A large class of other problems are solved by these same methods. \ In
particular, since the equipotentials are ellipsoids, the solution for the
uniformly charged line segment implicitly provides the solution for \emph{any}
charged \emph{conducting} prolate ellipsoid of revolution. \ This too is a
well-known fact
\cite{Green,Maxwell,KelvinTait,Routh,Durand53,Durand64,Smythe,Chandrasekhar,Rowley,Zangwill}%
. \ Thus the above results can be used to describe exactly the potentials and
electric fields for such ideal conductors.

Alternatively, the electrostatic results presented here may be used to
describe analogous Newtonian gravitational fields around massive focaloids.

Finally, since complicated circuits are often well-approximated by a sequence
of straight-line segments of various lengths, and since the magnetic field in
such situations is just the sum of the $\overrightarrow{B}$s\ for the
individual segments, my description for the magnetic field of a single segment
may help to understand $\overrightarrow{B}$\ for many circuits, even those for
which the field lines are very complex \cite{Lieberherr}.

\begin{acknowledgments}
I wish to thank Fulin Zuo for some useful exchanges about this subject. \ I
also thank Andrzej Veitia for bringing to my attention the information in
\cite{Puzzling}. \ Finally, I thank two anonymous reviewers for suggestions to
improve the manuscript.
\end{acknowledgments}


\begin{thebibliography}{99}
\bibitem {FootnoteA}In addition to $s=r_{a}+r_{b}$ and $L=b-a$, let
$t=r_{b}-r_{a}$. \ Then%
\[
\frac{b+r_{b}}{a+r_{a}}=\frac{2b+s+t}{2a+s-t}=\frac{a+b+L+s+t}{a+b-L+s-t}%
\]
Now comparing the two right triangles, with horizontal sides $a$ \& $b$,
hypotenueses $r_{a}$ \& $r_{b}$, and a common vertical side, gives the
relations
\begin{gather*}
st=r_{b}^{2}-r_{a}^{2}=b^{2}-a^{2}=\left(  a+b\right)  L\\
a+b+L+s+t=\frac{1}{L}\left(  s+L\right)  \left(  t+L\right) \\
a+b-L+s-t=\frac{1}{L}\left(  s-L\right)  \left(  t+L\right)
\end{gather*}
Therefore%
\[
\frac{b+r_{b}}{a+r_{a}}=\frac{s+L}{s-L}%
\]
\end{thebibliography}
\end{document}